\documentstyle[12pt]{article}
\begin{document}

\title{ON THE MATCHING CONDITIONS FOR THE  COLLAPSING CYLINDER}
\author{{L. Herrera$^1$\footnote{Postal address: Apartado 80793, Caracas
1080A, Venezuela.}
\footnote{e-mail: laherrera@movistar.net.ve}, M.A.H.
MacCallum$^{2}$\footnote{e-mail: m.a.h.maccallum@qmul.ac.uk} and N. O.
Santos$^{2,3,4}$\footnote{e-mail: N.O.Santos@qmul.ac.uk}}\\
\small{$^1$Escuela de F\'{\i}sica, Faculdad de Ciencias,}\\
\small{Universidad Central de Venezuela, Caracas, Venezuela.}\\
\small{$^2$School of Mathematical Sciences, Queen Mary, University
of London,}\\
\small{London, UK.}\\
\small{$^3$Universit\'e Pierre et Marie Curie - CNRS/FRE 2460,}\\
\small{LERMA/ERGA, Tour 22-12, 4\`eme \'etage, Bo\^{\i}te  142, 4 place
Jussieu,}\\
\small{75252 Paris Cedex 05, France.}\\
\small{$^4$Laborat\'orio Nacional de Computa\c{c}\~ao Cient\'{\i}fica,
25651-070 Petr\'opolis RJ, Brazil.}}
\maketitle

\begin{abstract}
We review  the matching conditions for a collapsing anisotropic
cylindrical perfect fluid, recently discussed in the literature
(2005 {\it Class. Quantum Grav.} {\bf 22} 2407). It is shown that
radial pressure vanishes on the surface of the cylinder, contrary
to what is asserted in that reference. The origin of this
discrepancy is to  be found in a mistake made in one step of the
calculations. Some comments about the relevance of this result
in relation to the momentum of Einstein--Rosen waves  are
presented.
\end{abstract}

\maketitle

\newpage

\section{Introduction}
In a recent paper \cite{HS}, using the Darmois conditions
\cite{Darmois}, the matching of a general collapsing anisotropic
cylindrical perfect fluid to the Einstein--Rosen (E--R) spacetime
\cite{Einstein} is studied in detail.

As the main result of that study, it appears that the radial pressure does
not vanish at the boundary surface of the matter distribution. The  physical
interpretation of this
 surface pressure is
justified through the continuity of the radial flux of momentum across
the boundary surface.

However, as we shall see in this note, such a result is incorrect. Indeed, a
careful review of the calculation process indicates a mistake in one of
the final steps. Once this mistake
is corrected, the final result is the vanishing of the radial pressure on
the boundary surface. This implies that any local pressure effect,
predicted by a local characterization of the
energy (momentum) flux of gravitational energy of the E--R waves, by means
of pseudo--tensors (see e.g. \cite{pseudo} and references therein), will
not be observed.

Let us show how this comes about.

\section{The collapsing perfect fluid cylinder and the Einstein--Rosen metric}

We consider a collapsing cylinder filled with anisotropic non-dissipative
fluid bounded by a cylindrical surface $\Sigma$.

We assume the general time dependent cylindrically symmetric metric
\begin{equation}
ds^2_-=-A^2(dt^2-dr^2)+B^2dz^2+C^2d\phi^2, \label{3}
\end{equation}
where $A$, $B$ and $C$ are functions of $t$ and $r$
and we
choose the fluid to be comoving in this coordinate system.

For the exterior vacuum spacetime of the cylindrical surface $\Sigma$ we
take the metric in Einstein-Rosen coordinates \cite{Einstein},
\begin{equation}
ds^2_+=-e^{2(\gamma-\psi)}(dT^2-dR^2)+e^{2\psi}dz^2+e^{-2\psi}R^2d\phi^2,
\label{26}
\end{equation}
where $\gamma$ and $\psi$ are functions of $T$ and $R$ satisfying  the field
equations $R_{\alpha\beta}=0$.

\section{Junction conditions and their consequences}

The  necessary and sufficient conditions for a smooth matching,
without a surface layer, are the Darmois junction conditions
\cite{Darmois,Bonnor1}, which require the continuity of the first
and second fundamental forms across the boundary surface.

The equations of $\Sigma$ may be written
\begin{eqnarray}
f_-=r-r_{\Sigma}=0, \label{30} \\
f_+=R-R_{\Sigma}(T)=0, \label{31}
\end{eqnarray}
where $f_-$ refers to the spacetime interior of $\Sigma$ and $f_+$ to the
spacetime exterior, and $r_{\Sigma}$ is a constant because $\Sigma$ is a
comoving surface forming the boundary of the fluid.

Using (\ref{30}) in (\ref{3}) we have for the metric on $\Sigma$
\begin{equation}
ds^2\stackrel{\Sigma}{=}-d\tau^2+B^2dz^2+C^2d\phi^2, \label{32}
\end{equation}
where we define the time coordinate $\tau$ only on $\Sigma$ by
\begin{equation}
d\tau\stackrel{\Sigma}{=}Adt, \label{33}
\end{equation}
and $\stackrel{\Sigma}{=}$ means that both sides of the equation are
evaluated on $\Sigma$.
We shall take $\xi^0=\tau$, $\xi^2=z$ and $\xi^3=\phi$ as the parameters on
$\Sigma$.

Then, the conditions on the
interior and exterior metrics imposed by the continuity of the first
fundamental form on $\Sigma$ are (see \cite{HS} for details)
\begin{eqnarray}
e^{\gamma-\psi}\left[1-\left(\frac{dR}{dT}\right)^2\right]^{1/2}dT\stackrel{
\Sigma}{=}d\tau, \label{35} \\
e^{\psi}\stackrel{\Sigma}{=}B, \label{36} \\
e^{-\psi}R\stackrel{\Sigma}{=}C. \label{37}
\end{eqnarray}

Next, differentiating (\ref{36}) and (\ref{37}) with respect to (\ref{33}) and
(\ref{35}) we obtain
\begin{eqnarray}
e^{2\psi-\gamma}\left[1-\left(\frac{\dot{R}}{\dot{T}}\right)^2\right]^{-1/2}
\frac{d\psi_{\Sigma}}{dT}\stackrel{\Sigma}{=}\frac{B_{,t}}{A},
\label{58} \\
e^{-\gamma}\left[1-\left(\frac{\dot{R}}{\dot{T}}\right)^2\right]^{-1/2}
\left(\frac{\dot{R}}{\dot{T}}-R\frac{d\psi_{\Sigma}}{dT}\right)\stackrel{\Sigma}{=}
\frac{C_{,t}}{A}. \label{59}
\end{eqnarray}
where
\begin{equation}
\frac{d\psi_{\Sigma}}{dT}
\stackrel{\Sigma}{=}\psi_{,T}+\psi_{,R}\frac{dR}{dT}, \label{63n}
\end{equation}
and the dot denotes the derivative with respect to $\tau$ in the surface.
Eqs.(\ref{58}) and (\ref{59}) are the corrected versions of  (36) and (37)
in \cite{HS}.

Then proceeding as in \cite{HS}, which involves the continuity of the
second fundamental form, we obtain (in agreement with Israel's result
\cite{Israel})
\begin{equation}
 P_{r}\stackrel{\Sigma}{=}0. \label{64}
\end{equation}
The result (\ref{64}) implies that  the flux of momentum of the
gravitational wave emerging from the cylinder will not produce any
observable local pressure.


\begin{thebibliography} {99}

\bibitem{HS} Herrera L and Santos N O 2005 {\it Class. Quantum Grav.} {\bf
22} 2407

\bibitem{Darmois}Darmois G 1927 {\it M\'{e}morial des Sciences
Math\'{e}matiques}, Gauthier-Villars, Paris, Fasc. 25

\bibitem{Einstein}Einstein A and Rosen N 1937 {\it J. Franklin Inst.} {\bf
223} 43

\bibitem{pseudo} Rosen N and Virbhadra K S 1993 {\it Gen. Rel. Grav.} {\bf
26} 429; Virbhadra K S 1995 {\it gr-qc/9509034}

\bibitem{Bonnor1}Bonnor W B and Vickers P A 1981 {\it Gen. Rel. Grav.} {\bf
13} 29

\bibitem{Israel} Israel W 1966 {\it Il Nuovo Cimento B} {\bf 56} 1

\end{thebibliography}
\end{document}